\documentclass[acmsmall, nonacm]{acmart}

\AtBeginDocument{%
  \providecommand\BibTeX{{%
    \normalfont B\kern-0.5em{\scshape i\kern-0.25em b}\kern-0.8em\TeX}}}

\usepackage{soul}
\begin{document}

\title{AI Product Security: A Primer for Developers}

\author{Ebenezer R.H.P. Isaac}
\orcid{0000-0003-0830-8862}
\affiliation{%
  \department{AI Technical Product Manager, Global AI Accelerator (GAIA)}
  \institution{Ericsson}
  \city{Chennai}
  \country{India}}
\email{ebenezer.isaac@ericsson.com}

\author{Jim Reno}
\affiliation{%
  \department{Distinguished Engineer, GAIA}
  \institution{Ericsson}
  \city{Santa Clara}
  \country{USA}}
\email{jim.reno@ericsson.com}

\renewcommand{\shortauthors}{E. Isaac and J. Reno}

\begin{abstract}
Not too long ago, AI security used to mean the research and practice of how AI can empower cybersecurity, that is, \textit{AI for security}. Ever since Ian Goodfellow and his team popularized adversarial attacks on machine learning, \textit{security for AI} became an important concern and also part of AI security. It is imperative to understand the threats to machine learning products and avoid common pitfalls in AI product development. This article is addressed to developers, designers, managers and researchers of AI software products. 
\end{abstract}




\maketitle

\section{Introduction}

Trustworthy AI is being explored by a number of jurisdictions around the world.  One example is the Ethics Guidelines for Trustworthy AI, from the High-Level Expert Group on AI set up by the European Commission. According to the
EC guidelines, trustworthy AI should be lawful, ethical and robust \cite{eu2019trustworthy}. The security of AI models is essential to addressing many of its
requirement areas,
which are becoming codified into laws and regulations, e.g., the EU AI Act \cite{euaiact}. As we continue to develop and rely on AI,
we must prioritize security and work 
to address the
challenges of AI safety.



The market for AI startups has exploded in recent years, with
many companies working on new and innovative applications.
Expertise in security is not a given among all those working in AI, which makes it essential to have a dedicated focus on it to ensure safe and secure AI systems.

The other day we came across this article titled ``Computer security checklist for non-security technology professionals.'' \cite{garrison2006computer} It is a short checklist grouped into three main activities: (1) Perform risk analysis, (2) Conduct vulnerability assessments, and (3) Education, procedures and policy. Though
published in 2006, these practices are still relevant today. AI Security, in today's world, goes a bit beyond that. From understanding the specific threats in machine learning (ML) to the crucial role of generic software product security, this article will take you on a journey through the ever-evolving landscape of AI security.

\section{AI-Specific Threats in ML}
\begin{figure}
  \centering
  \includegraphics[width=\linewidth]{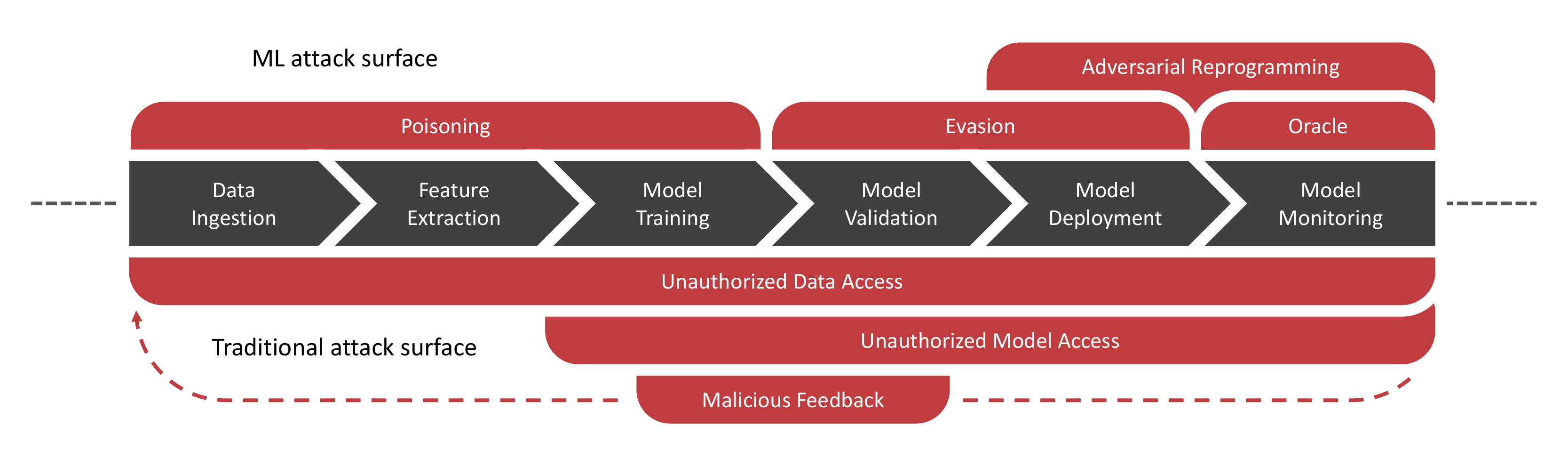}
  \caption{Taxonomy of threats of an ML product. Figure adapted from Hyrum Anderson \cite{anderson2021practical}}
  \label{fig:flow}
  \Description{A flow diagram highlighting various threats of an ML system in production}
\end{figure}

To address the complex challenges of AI security,
we need a holistic approach that covers all the aspects of the AI system, from data collection and preprocessing to deployment and monitoring. The taxonomy of
ML product attacks can be divided into two 
surfaces as shown in Fig~\ref{fig:flow}.

AI-specific attacks appear in the ML attack surface. These attacks are also collectively known as adversarial attacks.
\begin{itemize}
    \item Poisoning: modifying a benign training dataset 
    \item Evasion: give a malicious input to get an unexpected output
    \item Oracle: stealing information by probing the model
    \item Adversarial reprogramming: use the model for a task it is not intended to do
\end{itemize}

The traditional attack surface exploits vulnerabilities that can be found in any software product, regardless of its specific purpose or application.

\begin{itemize}
    \item Unauthorized access to data or the model can affect confidentiality, integrity, and availability of the system.
    \item Malicious feedback can negatively influence a model's development which may limit its ability to perform as expected.
\end{itemize}

Let us have a closer look at these attacks with examples and possible countermeasures.

\subsection{Poisoning}
Tampering with the training dataset by adding, deleting, changing, or reordering its contents can lead to erroneous learning and, ultimately, a model that generates incorrect inferences. Poisoning can violate one or more of availability, integrity, and confidentiality/privacy \cite{wang2022poisoning}. 

In 
some cases, poisoning
means adding malicious samples to the dataset,
affecting the learned concept to benefit the attacker.
An attacker
might poison the training dataset of a malware scanner to misclassify malware as benign code. A facial recognition system
might be tricked by introducing images of one
face
labeled with a different identity.
This false mapping might
used for identify theft or to warrant unauthorized access, e.g., unlocking a phone or
restricted areas.

In a classification model, an attacker can inject multiple copies of seemingly benign samples into the training dataset. If more of
one class is supplied, then the integrity of the output is affected by
changing the convergence -- the point where the model is said to have learned its target concept.
Injecting an abnormally high volume might increase training computational requirements, stall the pipeline, and reduce system availability.

Consider a linear regression model which forecasts financial information.  If the attacker can poison the training data,
the subsequently corrupted model may give him some
advantage.  For example, he might affect the predicted price of a stock in order to facilitate securities fraud.

Risk of poisoning is greater in products where there is limited control on data collection, such as crowd-sourcing or federated learning. The following steps may help mitigate the risks.

\begin{itemize}
    \item Assess
    the trustworthiness of participants.
    Ensure that only verified members can provide inputs 
    to the dataset \cite{papadopoulos2021privacy}. Verification 
    can involve authorization of members contributing to the dataset and authenticity of messages before they are used to train an ML model.
    \item Assess load prior to training and preprocessing to ensure
    the pipeline
    won't stall. Include exceptions when the data exceeds the expected distribution. E.g., depending on the use case,
    a subset
    can be preferred
    over the entirety of the input data for training.
    \item Assess class-wise distribution to detect data drifts. Outliers, particularly if they consistently come from a small subset of participants, could indicate an attack. Depending on the use case, this method may not
    distinguish natural drift from a poisoning attack. Nevertheless, it can alert the model operator to initiate an investigation.
    \item When crowd-sourcing, use randomly selected subsets taken from a very large community for the training and test datasets.  This approach increases the amount of work the attacker must do in order to have enough poisoned data points included in the model.
\end{itemize}

It is worth noting that these methods are not a panacea and need to be combined with other techniques like data validation, data sanitizing and model validation to
protect against poisoning.

\subsection{Evasion}
Evasion attacks
manipulate input data in a way that causes a
model to make incorrect
decisions.
With slight, ideally unnoticeable, input modifications, the attacker may be able to cause an output that is in 
her favor. The modified sample is called an \textit{adversarial example}. Vulnerabilities to evasion attacks are common in image classification systems (usually neural networks). Goodfellow et al. \cite{goodfellow2015explaining} show
an image of a panda
perturbed to make GoogLeNet misclassify it as a gibbon
with 99.3\% confidence. Both images before and after perturbation look identical to the human eye.
The perturbation is done only on the sample fed to the model,
without altering model parameters/weights.

An evasion attack is a direct violation of integrity,
but can also violate confidentiality (if the model is used to authorize access) or availability (e.g., a virus
that bypasses malware checks by warping its signature). A detailed outlook on
evasion attacks has been studied in reference \cite{sagar2020applications}.

Specific risk mitigation for evasion attacks is influenced by the application that uses the ML model, such as malware detection, phishing detection, internet of things (IoT), smart grids, etc. Nevertheless, studying the threat surface can help to identify possible controls to reduce this risk. Start by answering the following questions:

\begin{itemize}
    \item Where in the pipeline can a possible perturbation occur?
    \item What is the data source for the example used for inference and validation?
    \item Is the pathway from the source to the inference engine hardened?
\end{itemize}

For instance, in an image classification scenario, perturbations can occur in the image samples.
In face detection, the data source is the camera. Ensure that the data flow from camera
$\rightarrow$
storage $\rightarrow$  inference
is secure.
Where you have minimal control over the data source, you may apply certain transformations on the data \cite{yuan2019adversarial}. Defensive distillation is a
way to
strengthen a learning algorithm against potential adversarial attacks by adding more adaptability to its classification process. It involves training one model to predict the output probabilities of another model that was trained on a previous, standard dataset, focusing on overall accuracy \cite{papernot2016distillation}. However, defensive distillation is not robust against poisoning,
so adequate protections
against poisoning should also be in place. 

\subsection{Oracle}
An oracle attack in ML, or
model extraction/stealing attack \cite{jagielski2020high}, involves an attacker attempting to extract the internal parameters or architecture of a model
by querying it
to infer its decision boundaries. The goal
is to recreate a copy of the model and potentially use it for malicious purposes, such as stealing sensitive information or
intellectual property. This type of attack can happen when the attacker has access to the model's predictions, but not the training data or the model's parameters. The attacker can construct a specific
set of queries to the model to infer the underlying logic or data. The information gained from an oracle attack can also be used to facilitate other types of attacks. Two interesting types of oracle attacks are membership inference attacks and model inversion attacks. 

In a membership inference attack \cite{carlini2022membership}, an attacker aims to determine whether a specific individual's data was used to train a machine learning model. The attacker makes use of the model's predictions
as an "oracle", sometimes correlated with publicly available data, to infer whether the individual's data was used in the training process. This can reveal sensitive information,
compromising the privacy of the individual. For example, an attacker could use a model that has been trained on
medical records to infer whether a specific individual's medical data was
in the training set.  The attacker
submits queries to the model with various combinations of attributes of the individual, and observes the model's predictions. If the
predictions match the individual's known attributes, the attacker can infer that the individual's data was
present.
A case study of this attack and its defence
is discussed in \cite{mccarthy2023defending}.

A model inversion attack \cite{fredrikson2015model}, on the other hand, is an attack where an adversary tries to reverse engineer the training data that was used to train a machine learning model. By doing this, the adversary can uncover sensitive information from the training data, such as the identity of people or the exact geographic locations that were used to train the model.

Oracle attacks exploit the availability of model. One simple
defense is to limit
access to the model, similar to
defending against denial of service (DoS) attacks. The success
of an oracle attack depends on the number of queries the attacker can submit.
Throttling query access can slow or prevent an attack.  Frequent model updates can also help, by changing the model before an attacker
can submit a sufficient number of queries. Note that when
performing transfer learning on a pretrained model that is already
public (such as ResNet50 trained on ImageNet \cite{he2016deep}) or used by another entity, then this vulnerability can still exist even though you curb the availability of the model.
Here the attacker can gain access to the base pretrained model and have sufficient time and access to
create a set of curated adversarial examples, and compare them to the target system. 
This can be done even if the attacker only has a few opportunities to test instances and observe the output. 


\subsection{Adversarial Reprogramming}
\textit{Untargeted} adversarial attacks aim to affect the performance of a model without a specific target output.
\textit{Targeted} attacks
create an adversarial perturbation to force a particular output for a given input.
Adversarial reprogramming, as demonstrated by Goodfellow and his team \cite{elsayed2019adversarial}, goes beyond this
by considering the ability of an attacker to alter the intended function of a machine learning model.
They were able to redirect an ImageNet object classification model to perform a separate counting and classification task. This highlights the potential for malicious actors to manipulate models for their own purposes ranging from the misuse or theft of compute resources to secret message passing treating systems as spies.

Just like evasion and oracle, this attack does not require
changing the model weights.
Once enough knowledge is gained from
inputs and outputs of the system, for instance through an oracle attack, an attacker can craft a
program to create an adversarial example.
The model will map that example to the output chosen by the attacker.
The initial assumption for this attack is that it requires a white-box knowledge of the model, i.e., the structure and weights of the model are known to the attacker. However, it was proved later that this attack can be done even with a black-box model wherein the weights and structure of the network are unknown to the attacker \cite{tsai2020transfer}. 


 Protection of the model's availability and parameters can help in controlling the vulnerability of an adversarial reprogramming attack. The use of adversarial training methods like defensive distillation do not eliminate its vulnerability to reprogramming (although it might increase the compute required). Other ways to reduce the risk of such an attack are regularly updating the model to ensure that it remains resilient against new types of attacks, and monitoring the model's performance in real-world scenarios to detect any unusual behavior.

\subsection{Traditional Attacks}

Data flows in the pipeline through multiple components from ingestion to monitoring. An attacker can target the traffic flow to access confidential information or even participate in a man-in-the-middle attack. A model also undergoes various states as it passes through the pipeline. Once trained, it holds the essence of the data in the form of weights. In production, the model resides in a model store, and is served during inference. The model can be
attacked at any point in these state transitions.
Some pipelines have feedback mechanisms to support
model retraining.
Of course, if feedback from many sources is
used,
then a few malicious values need not
inhibit the training. However, if the feedback flow itself is compromised, then an attacker
can inject/malform any data.

Some techniques under research for addressing the traditional attack surface of an AI system
include secure generation of the training set, training set obfuscation, and
securely acquiring the sample at inference \cite{khalid2018security}. By focusing on the security of the overall product, not just the machine learning component, organizations can protect their assets and stakeholders. Hardening the ML pipeline can defend against traditional attacks and possibly some AI-specific attacks.
It also helps to maintain the overall integrity and confidentiality of
data, and to comply with various regulations and industry standards.
Security controls include (but are not limited to) the following:

\subsubsection{Constraining listening services and bindings}
Binding a service
connects it to a specific endpoint or location, such as a network address or
port. On a default setup of an end-to-end system, one
Often finds many accessible services. Even if these services require secure credentials,
it is a good practice to disable
access to those that are not required outside the product boundary. By restricting access to  the services and bindings,
it becomes more difficult for an attacker to exploit vulnerabilities within the system. 
Constraining listening services and bindings can also help to improve performance and reduce the overall complexity of the AI system.

\subsubsection{Impose strict access control mechanisms}
All software should have adequate authentication and authorization. The principle of least privilege (PoLP) is a
policy
ensuring users are only granted
enough access
to perform their tasks and no more \cite{steiner2018hardening}. It ensures that only authorized individuals or entities
can access
or change the system, reducing the risk of unauthorized access, data breaches and other malicious activities. This is especially important for AI systems where seemingly innocuous privileges (e.g., API query access) can enable certain attacks (e.g., oracle or inference attacks).

\subsubsection{Traffic and data protection}
Traffic protection means measures implemented to secure incoming and outgoing communication, usually at the IP level.
Not all flows
passing between product components are equal. For instance, there should be separate flows for each plane of traffic \cite{jonesRFC3871}. These planes may include Operation and Maintenance (O\&M), data plane, control plane, and user plane traffic. The traffic should also be secured by an up-to-date version of transport layer security (TLS) \footnote{As of the time of writing, 1.3 is the preferred version of TLS \cite{rescorlaRFC8446}}. Data at rest should also be encrypted since a
data leak can expose sensitive information or
help an attacker
subvert or invert the model.  Zero-trust principles require that communication channels not only be encrypted, but also authenticated in both directions. Data protection includes the data pipeline all the way back to the source.

\subsubsection{Periodic vulnerability analysis}
Vulnerability assessment is the process of identifying, quantifying, and prioritizing
the vulnerabilities in a system.
This can be done manually or
with automated tools such as grype \cite{grype} and trivy \cite{trivy}. 
One
key
step of a vulnerability assessment is identifying Common Vulnerabilities and Exposures (CVEs) \cite{cve} -- a standardized way of describing a vulnerability or security weakness.
Each has a unique identifier and includes information such as a brief description of the vulnerability, the affected software or system, and its severity. Vulnerabilities are found frequently, with several new ones identified every day. Fixing them might seem like a never-ending process, but it is important to prioritize on the most critical ones and address them before the product hits the market, and then do periodic fixes. For development teams under tight deadlines this means
a careful tradeoff between time, cost and risk.

Full knowledge of vulnerability management in software product security
requires a course of its own. Fortunately there are resources to help.
The MITRE ATT\&CK \cite{mitreattack} provides a comprehensive understanding of the tactics and techniques used by attackers.
The NIST Cybersecurity Framework \cite{nistCSF} provides
a structured approach to managing their cybersecurity risks. The recently announced NIST AI Risk Management Framework Playbook \cite{nistAI} builds on the NIST Cybersecurity Framework and provides organizations with specific guidance on how to manage the unique risks posed by AI systems.   OWASP (Open Web Application Security Project) 
keeps an up-to-date summary of the top 10 vulnerabilities that pertain to web applications \cite{owasp10}. 
Counterfit \cite{counterfit} is another open-source tool released by Microsoft for automating security testing of AI systems. It helps organizations perform AI security risk assessments to ensure their algorithms are robust, reliable, and trustworthy.
By using these tools, organizations can develop a comprehensive cybersecurity strategy that accounts for the complex threat landscape and helps them effectively manage their risk.


\subsubsection{Secure coding practices}
Secure coding is a method of developing software to prevent the accidental introduction of security vulnerabilities.
It includes security code reviews, 
security education, and use of automatic code analysis tools. One
secure coding framework
is the Software Engineering Institute (SEI) Computer Emergency Response Team (CERT) Coding Standard \cite{seicert}. CERT has standards sets for C, C++, Android, Oracle Java, Perl. As of now, CERT does not have a standard set for Python. Bandit \cite{bandit} is a Python linting tool
that helps developers identify and prevent security-related code issues in
Python applications. Bandit checks the code against a set of predefined security rules and generates a report indicating the potential security issues.

\subsubsection{Event logging}
Logging provides a record of events and activities
within the system. It can provide
evidence in the event of a security incident
by showing what actions were taken and when they occurred. In addition to security assessments, logs can be used for a variety of purposes, including debugging, troubleshooting, auditing, and monitoring.
The system must be auditable: Ideally,
every user-initiated action should be
traceable back to the user
that initiated the action.
Logs that show a sequence of activities that lead to the source of the action are called audit trails. 

In AI systems,
logs can include the queries made to the model
(including the parameters),
the model version, and the inference data feed.
Logging involves tradeoffs. When every possible event is logged, the system is said to be more secure, but more compute and storage is required.
In ML, depending on the querying frequency and the inference data size, the storage requirement for logging may grow exponentially with respect to the number of interoperable components. So, one should be mindful of the items
logged and how long
logs should be kept. One possibility is to
make such parameters
configurable with adequate documentation (such as a Security/Privacy User Guide).
The user of the AI product is then aware and can make an informed decision on what to be logged.
Also keep sensitive data out of logs, and control log access through PoLP.


\subsubsection{Ensure runtime environment security}
Even if the
model is developed with
security principles in mind, if
deployed in an insecure environment,
the model and its inputs and outputs can be attacked.
This
facilitates targeting the system with the entire spectrum of adversarial attacks. Hence, employ measures that
harden the runtime environment.
The security measures mentioned in this traditional attack section are equally applicable to the runtime environment. Other measures include encrypting storage, monitoring interfaces, and periodic security patch updates. 

\subsubsection{Secure the supply chain and development pipeline}
Much of today's AI code is open source, making it easy for vulnerabilities to be identified and exploited. Patches
come out periodically.
Fixing those vulnerabilities is a continuous process. 
Securing the software development supply chain involves implementing
measures to ensure the integrity and security of the components and services used throughout the software development life cycle.
Supply chain security includes:

\begin{itemize}
    \item Verifying the authenticity of the components and services used, such as open-source libraries, by checking their signatures and digital certificates.
    \item Using a private repository or registry to store and manage internal components, reducing the risk of using untrusted or malicious components.
    \item Regularly monitoring and updating dependencies, including open-source libraries.
    \item Implementing security practices such as code signing, secure packaging, and continuous integration and deployment (CI/CD) to prevent tampering and maintain the integrity of the software. The CI/CD system itself should
    be secured appropriately as a production system.
    \item Training employees to recognize and avoid potential supply chain threats, and implement security policies and procedures to reduce the risk of a successful attack.
\end{itemize}

More detail on best practices for software supply chain security is provided by the Cloud Native Computing Foundation (CNCF) \cite{cncfsupplychain}


\section{Security Pitfalls in AI Product Development}
Security is an important aspect of
AI products, yet it is often overlooked during
development.
Many AI
developers focus primarily on building the technology and achieving high performance, without considering the potential security risks and vulnerabilities. This lack of security focus can result in AI products that are vulnerable
and may not protect the data and information they process.
 
\subsection{Lack of security expertise in your team}
Security is likely to be the last thing on the mind of a Data Scientist, Machine Learning Engineer, or AI Product Developer. Unless there is a dedicated security expert in your team, you are bound to overlook some security or privacy flaw that may creep into development. You cannot expect the entire team to undergo the rigorous security training that is expected of a security professional. Nevertheless, a minimal security training program should exist for all members, and a security expert
made available to
help with the details.
For instance, a Data Scientist may not be expected to know the current TLS standards to protect the data in transit between one component to another. 

\subsection{Missing
security requirements at the start of a project engagement}
When security requirements come late in the project, it can create complications
for development and program management: Who takes ownership of the activity? Who has the 
appropriate expertise?
What overhead is added to the project? Even if
the project itself has little to do with security, an early security discussion will avoid unpleasant surprises.

\subsection{Assuming security aspects will be taken care of by another team.}
Within a dedicated product development unit, the roles
are often well-defined.
There might be a dedicated team to handle security-related issues. Program management expects a representative from this team to participate in
requirements discussions to close security gaps. However, when working with cross-functional teams, responsibilities aren't always clear.
It is vital in such situations to set clear expectations as to
who handles the security function rather than assuming "it will be taken care of by another team". It is best if \textit{every} team member has knowledge and involvement in security, but at a minimum, those responsible should be clearly understood and documented.

\subsection{Delaying security and privacy compliance activity towards the end}
Envision this scenario: your team has put in all their effort to polish a product based on their priority functional requirements. The product then goes through a security assessment. The report uncovers
some
critical vulnerabilities, and component integration has introduced some traffic threats. Your models
aren't encrypted and there are some packages that you should not even use in production. In the worst case, this can lead to an overhaul of the entire product design. Solution: \textit{shift-left security}. This practice involves bringing the best practices of security earlier in the product %
design 
deveopment process - ideally from the very beginning - to avoid such issues. 

\subsection{Using features without evaluating their sensitivity.}
Let's say you showcase a successful proof of concept (PoC), everyone gets excited, and the sales team wants to show it to the customers as soon as possible.
In the assessment, you notice
sensitive fields that
shouldn;t be in the model, such as gender, region, and age. Once you remove these fields and retrain,
model performance takes a hit – far below the promised value. The product ultimately fails to meet its requirements.
This
might have been averted by conducting a privacy impact assessment for all fields before they are considered features (or used to derive features) for fitting the model.

This pitfall does not necessarily have to be ML model-related. In telecom, subscriber data is highly sensitive. Examples include phone number (MSISDN) and IMSI. When these items are not specifically needed by the model, they should not be collected, or should be masked or removed from the training data.  If they are required, consider using privacy-preserving mechanisms such as de-identification or anonymization. For instance, not all network operators need to access this data when using a dashboard for anomaly detection of cell traffic.


\section{Closing thought}
Security of any kind
is not just a checklist – it is a process. AI security is no different in this regard. It is crucial that security be integrated into the development process from the start and not as an afterthought. 
Work on the following to help avoid the aforementioned pitfalls.
\begin{itemize}
    \item Understand the implications of AI
    security and address those gaps in a ML pipeline
    \item A minimal security training programme for non-security professionals
    \item Discuss security requirements and responsibilities as early as possible; preferably before the start of the project engagement
    \item Shift-left security and security by design
    \item Assess the privacy impact of your data before committing it to a model
    \item Be wary of AI-specific attacks like poisoning, evasion and model/data access
    \item Periodic security audits and testing against latest security vulnerabilities.
\end{itemize}
The goal of AI is not just to prove a possibility 
but to build a something that provides value to us and our customers. That means that the 'mundane' aspects are just as important as the data science.

\begin{acks}
To Amit Sharma, Head of AI Hub India 2 at GAIA, Kalpana Angamuthu, Senior Manager of Data Science at GAIA, and Michael Liljenstam, Principal Researcher at Ericsson Research, for reviewing this work and sharing their valuable suggestions.
\end{acks}

\bibliographystyle{ACM-Reference-Format}
\bibliography{AI-Security}




\end{document}